\documentclass{elsarticle}
\usepackage{graphicx}
\usepackage{amsmath}
\usepackage{bm}
\usepackage{amsbsy}
\usepackage{subfigure}
\usepackage{CJKutf8}
\usepackage{epstopdf}
\usepackage{epsf,epsfig,graphics}
\usepackage{float}
\usepackage{makeidx}
\usepackage{latexsym}
\usepackage{amssymb}
\usepackage{amsfonts}
\usepackage{mathrsfs}
\usepackage{dcolumn}
\usepackage{amsmath}
\usepackage{color}
\usepackage{lineno,hyperref}
\modulolinenumbers[5]
\journal{Journal of \LaTeX\ Templates}

\usepackage{morefloats}
\DeclareMathAlphabet\mathbfcal{OMS}{cmsy}{b}{n}

\begin{document}
\begin{frontmatter}
\title{Non-equilibrium Dynamics and Phase Transitions in Potts model and Interacting Ehrenfest urn model}

\author{Chi-Ho Cheng$^a$\footnote{phcch@cc.ncue.edu.tw}
 and Pik-Yin Lai$^{b,c}$\footnote{pylai@phy.ncu.edu.tw}
}
\address{$^a$Dept. of Physics, National Changhua University of Education, Changhua 500, Taiwan, R.O.C.}
\address{$^b$Dept. of Physics and Center
for Complex Systems, National Central University, Chung-Li District, Taoyuan City 320, Taiwan, R.O.C.}
\address{$^c$Physics Division, National Center for Theoretical Sciences, Taipei 10617, Taiwan, R.O.C.}

\begin{abstract}
We show that the recently proposed interacting Ehrenfest $M$-urn model at equilibrium[Phys. Rev. E  101 (2020) 012123] can be exactly mapped to a mean-field $M$-state Potts model. By exploiting this correspondence, we show that the $M$-state Potts model with $M\geq 3$, with transition rates motivated by the non-equilibrium urn model, can exhibit rich non-equilibrium spin dynamics such as non-equilibrium steady states and non-equilibrium periodic states.
Monte Carlo simulations of the 3-state Potts model are performed to demonstrate explicitly the first-order transitions for the equilibrium and non-equilibrium steady states, as well as the far-from-equilibrium periodic states.
\end{abstract} 

\end{frontmatter}

\section{Introduction}
The $M$-state Potts model\cite{Potts} is a generalization
of the classic Ising model\cite{Ising} to $M$ spin components.
 It is known that the
Potts model is related to several paradigm systems
in statistical physics \cite{PottsWu}, in particular for phase transitions in equilibrium statistical mechanics and its critical behavior has also
been shown to be richer than that of
the Ising model. For example, it was  shown\cite{KF1969} that the problem of the bond percolation
can be formulated in terms of the $M=1$ Potts model which was extended further to the site percolation problem\cite{Giri}. 
The Potts model also finds its applications in a variety of systems related to graphs or networks, such as the chromatic number problem of graph coloring which is closely related to the ground state of the anti-ferromagnetic Potts model\cite{PottsWu, Baxter}. Other applications include the graph partition problem\cite{graph87,graph88}, the network-community detection\cite{Pottscommunity}, and even the cell-differentiation model in developmental biology\cite{Glazier}.
Although the equilibrium Potts model and its phase transition behavior have been rather thoroughly investigated, its non-equilibrium dynamics are much less studied as compared to the non-equilibrium Ising spin dynamics.

On the other hand, the less well-known Ehrenfest urn model\cite{urn1907,karlin1965,kelly1979}, proposed more than a decade before the Ising model aimed to explain or illustrate the Second Law of thermodynamics, was one of the earliest models for non-equilibrium statistical physics.
The original Ehrenfest two-urn model\cite{urn1907} is a simple model with no interaction that is tractable and can illustrate the conceptual foundation of statistical mechanics for the relaxation towards equilibrium. Recently, the two-urn Ehrenfest model was extended to include particle interactions inside an urn\cite{cheng17} in which particles can interact with all other particles inside the same urn, but particles belonging to different urns do not interact. In addition,   
a  jumping rate (asymmetric in general) from one urn to another can also be introduced, which is independent of the particle interaction. The system was shown to exhibit interesting phase transitions and the Poincar\'e cycle together with the relaxation times can be calculated\cite{cheng17}.
 The interacting Ehrenfest two-urn model was subsequently extended to $M$ urns ($M\geq2$) for the equilibrium case in which detailed balance can be achieved. The equilibrium phase behavior is rich and can be investigated in detail\cite{cheng20}. Similar to the two-urn case\cite{cheng17}, $N$ particles are distributed into the $M$ urns. Pairwise all-to-all interaction is  introduced only for particles in the same urn and particles in different urns
do not interact.
 As the inter-particle interaction strength is varied,  phases of different levels of non-uniformity emerge and their stabilities are calculated analytically. In particular, the coexistence of locally stable uniform and non-uniform phases connected by first-order transition occurs\cite{cheng20}. The phase transition threshold and energy barrier can be derived exactly together with the phase diagram obtained analytically. Analytic and exact results are derived for the condition for the emergence of coexisting uniform or non-uniform phases and the associated first-order phase transition and energy barrier.

Very recently, the $M$-urn model with intra-urn interactions has been extended to the non-equilibrium scenario by arranging the urns in a ring and introducing asymmetric clockwise and counter-clockwise jumping rates\cite{cheng21}. It was demonstrated that the system can exhibit two distinct non-equilibrium steady states (NESS) of uniform and non-uniform particle distributions with the associated non-equilibrium thermodynamic laws revealed.
In addition,  a first-order non-equilibrium phase transition occurs\cite{cheng21} between these two NESSs as the inter-particle attraction varies. The phase boundaries, the NESS particle distributions near the NESSs and the associated particle fluxes, average urn population fractions, and the relaxational dynamics to the NESSs were obtained analytically and verified numerically\cite{cheng21}.

As for spin systems, non-equilibrium spin dynamics usually result from the relaxation of a quenched external field or external time-depending driving. 
Periodic oscillations in magnetization can be easily obtained with an external time-dependent oscillatory drive, but autonomous oscillations in interacting spin systems are hardly reported. 

In this paper, we show and demonstrate that the above seemingly independent Ehrenfest $M$-urn model and  $M$-state Potts model are in fact closely related and can be mapped to each other. Furthermore, the non-equilibrium dynamics of these systems reveal non-trivial NESSs and non-equilibrium period states (NEPS), which may serve as a paradigm statistical physics system to investigate systems with different levels away from equilibrium.
Sec. 2 gives a summary of the $M$-urn model and its phase transition behavior at equilibrium, together with the non-equilibrium model of the $M$ urns arranged on a ring. The equivalence of the $M$-urn model with intra-urn interaction at equilibrium and the mean-field $M$-state Potts model is established in Sec. 3.
The Potts spin dynamics that correspond to the non-equilibrium $M$-urn model on a ring are derived in Sec. 4. These theoretical results are verified by the Monte Carlo simulations of the 3-state Potts model in Sec. 5. Sec. 6 gives the concluding remarks and future outlooks.

\section{Summary of the equilibrium and non-equilibrium Ehrenfest urn models with interactions}
In this section, we give a summary of the equilibrium and non-equilibrium Ehrenfest urn models and their associated properties. Details of these models can be found in Refs. \cite{cheng20} and \cite{cheng21}. For the $M$-urn model, a total of $N$ particles are placed inside $M$ urns and the number of particles in the $\alpha^{th}$ urn is denoted by $n_\alpha$, with $\alpha=0,1,\cdots,M-1$. Particles from  different urns
do not interact whereas any pair of particles in the same urn interact with an energy specified by the urn.
Since the total particle number $N$ is fixed,
the constraint that $n_0 + n_2 +\cdots+ n_{M-1} = N$ is always satisfied. The energy or Hamiltonian of the interacting particles in the urns is given by
\begin{equation}
\beta{\cal H}=\frac{1}{2N}\sum_{\alpha=1}^M g_\alpha n_\alpha(n_\alpha-1),\label{Ham}
\end{equation}
where $\beta\equiv 1/(k_B T)$ is the inverse temperature and $g_\alpha$ is the pair-wise interaction (in the unit of $k_B T$) of the particles inside the $\alpha^{th}$ urn.
It is found that for homogeneous coupling with $g_\alpha=g$, an equilibrium first-order phase transition occurs between the uniform and the first non-uniform states as $g$ varies and the first-order transition occurs at $g=g_t$ which is given by\cite{cheng20} 
\begin{eqnarray}
g_t&=&-\frac{2 (M-1)}{M-2} \ln  (M-1).\label{gt01}
\end{eqnarray}
To quantify how non-uniform the state is, one can  define
 \begin{equation}
    \Psi=\sqrt{\frac{1}{M(M-1)}\sum_{\alpha\neq \alpha'}(x_\alpha-x_{\alpha'})^2}\label{psi}
\end{equation}
as the non-uniformity of the state,  where $ x_\alpha\equiv\frac{n_\alpha}{N}$ is the population fraction in the $\alpha^{th}$ urn. $\Psi$ can also serve as an order parameter for the phase transition: $\Psi= 0$ for the uniform (disordered) state and $\Psi>0$ for the non-uniform (order) state.

For the non-equilibrium urn model, particles undergo transitions among the urns, in general one can consider $M$ urns are placed on a network with possible transitions between the urns represented by the edges.  The  transition probability of a particle in the $\alpha^{th}$ urn  jumps to
the $\alpha'^{th}$ urn is denoted by $T_{\alpha\to\alpha'}$. As in Ref. \cite{cheng21}, here we focus on urns placed on a one-dimensional ring for convenience to investigate the cyclic particle fluxes.  In addition,  a direct jumping rate is further introduced such that the probability
of anticlockwise (clockwise) direction is $p$ ($q$). The condition $p+q=1$ is imposed which changes only the time scale. For the 3-urn system, it has been shown\cite{cheng21} that two distinct NESSs of uniform and non-uniform particle distributions, separated by a first-order non-equilibrium phase transition, occur as the interparticle attraction varies,  with the presence of a coexistence regime for smaller values of $|p-q|$.

 NESS corresponds to the situation that an urn receives several particles per unit of time from the upstream neighboring urn and sends out an equal number of particles per unit of time to the downstream neighboring urn such that the number of particles in this urn remains basically unchanged (subject to stochastic fluctuations). If the inter-particle attraction is not so strong, the particle numbers in each urn are the same on average resulting in a uniform NESS. On the other hand,  when the inter-particle attraction becomes strong enough, an urn can have more particles and maintain the population on average because of stronger attraction resulting in a non-uniform NESS.
 Remarkably, under some parameter regimes in which the driving is strong and the particle attraction is also significant, the fast flow of particles from upstream will be accumulated on average in an urn for some duration before their particles can be released to the downstream urn to deplete the population below average. This can lead to a stable periodic oscillation of the population in an urn and the particle flux in the ring, resulting in the NEPS\cite{cheng23}.
 
\section{Equivalence of the  Ehrenfest Urns Hamiltonian to the Potts model}
\begin{figure}[H]
\centering
  \subfigure[]{\includegraphics*[width=.48\columnwidth]{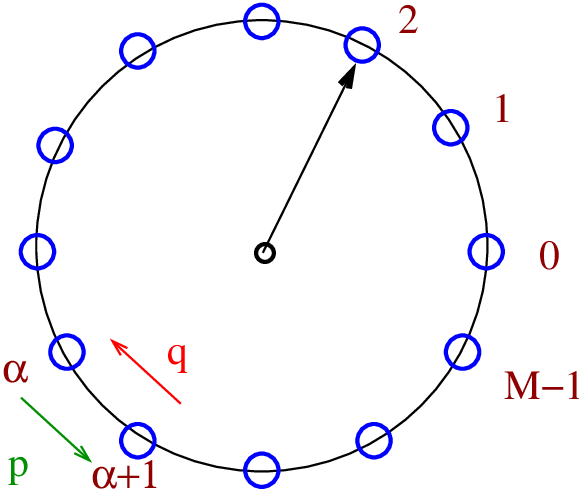}} 
    \subfigure[]{\includegraphics*[width=.48\columnwidth]{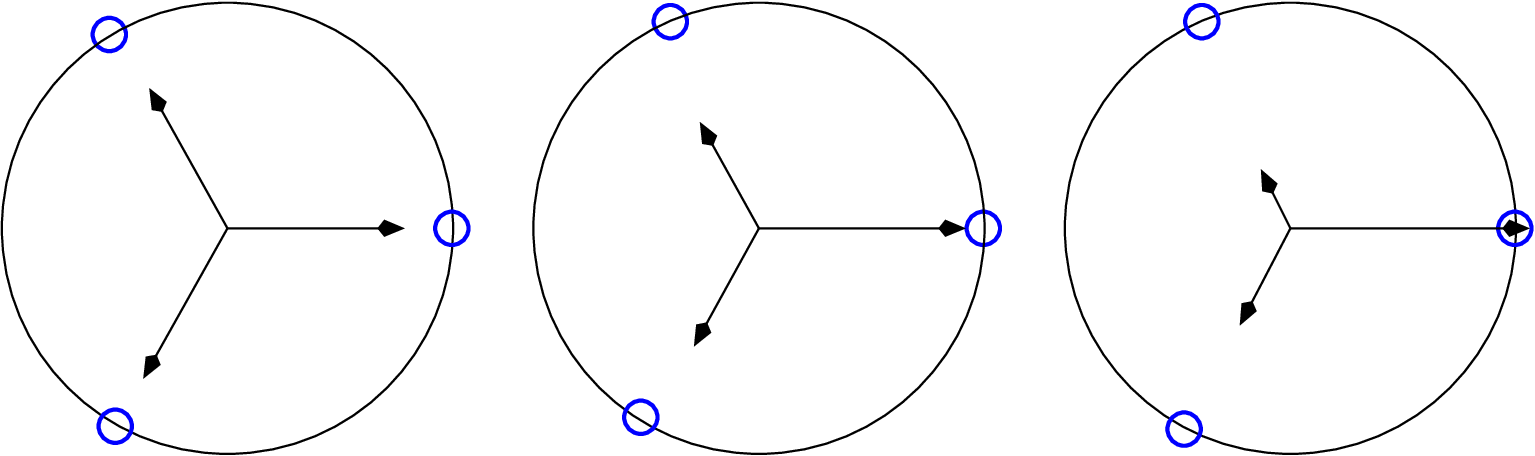}} 
  \caption{Schematic picture showing the equivalence of the interacting Ehrenfest $M$-urn model on a ring and the $M$-state Potts spin with discrete $Z_M$ symmetry. (a) $M$ urns (denoted by the open circles) are placed on a ring. The $M$ open circles also represent the  $M$ discrete Potts spin directions on the unit circle. The non-equilibrium dynamics are introduced by the direct counter-clockwise and clockwise jumping rates $p$ and $q$ respectively.  (b) Schematic pictures showing the average magnetization in each direction of the 3-state Potts model under uniform/disordered state (left), non-uniform/ordered state (middle), and non-uniform non-equilibrium steady state (right).}
  \label{schematic}
  \end{figure}

The $M$-urn model at equilibrium can be mapped to the mean-field $M$-state Potts model with the $i^{th}$  spin (particle) denoted by $\sigma_i$ which can take values  $\sigma_i=0,1,\cdots,M-1$. We shall use indices $i,j,k,\cdots$ for the spin/particle label and the Greek alphabet $\alpha=0,1,\cdots,M-1$ for the urn or Potts state label. Note that the population fraction in the $\alpha^{th}$ urn is given by
\begin{equation}
    x_\alpha\equiv\frac{n_\alpha}{N}=\frac{1}{N}\sum_{i=1}^N\delta_{\sigma_i\alpha},\label{xa}
\end{equation}
which can be viewed as the mean ``magnetization" of the Potts spin along the $\alpha$-direction (see Fig. \ref{schematic}a). 
The interacting urn Hamiltonian in Eq. (\ref{Ham}) can be expressed as an interacting Potts spin Hamiltonian using Eq. (\ref{xa}): 
\begin{eqnarray}
    \beta{\cal H}&=&
    \frac{1}{2N}\sum_{\alpha=0}^{M-1} g_\alpha n_\alpha^2- \frac{1}{2N}\sum_{\alpha=0}^{M-1} g_\alpha n_\alpha=\frac{N}{2}\sum_{\alpha=0}^{M-1} g_\alpha x_\alpha^2-\frac{1}{2} g_\alpha x_\alpha  \nonumber \\
    &=&\frac{1}{2N}\sum_{i,j}^N\sum_{\alpha=0}^{M-1} g_\alpha\delta_{\sigma_i\alpha}\delta_{\sigma_j\alpha}-\frac{1}{2N}\sum_{i,j}^N\sum_{\alpha=0}^{M-1} g_\alpha\delta_{\sigma_i\alpha} \nonumber \\
    &=&\frac{1}{2N}\sum_{i,j}^Ng_{\sigma_i}\sum_{\alpha=0}^{M-1}\delta_{\sigma_i\alpha}\delta_{\sigma_j\alpha}-\frac{1}{2N}\sum_{i,j}^Ng_{\sigma_i}\sum_{\alpha=0}^{M-1}\delta_{\sigma_i\alpha} \nonumber \\
   &=& \frac{1}{N}\sum_{i<j}^Ng_{\sigma_i}\delta_{\sigma_i\sigma_j},\label{PottsHam}\end{eqnarray}
which is a Potts model with heterogeneous couplings that depend on the spin directions. In general, the energy change due to the flipping of the $k^{th}$  spin is given by
\begin{equation}
    \beta\Delta {\cal H}(\sigma_k\to\sigma_k ')=g_{\sigma_k '}x_{\sigma_k '}-g_{\sigma_k} x_{\sigma_k}.\label{DelH}
\end{equation} In particular, for homogeneous coupling with $g_\alpha=g$ for all $\alpha$, Eq. (\ref{PottsHam}) is the mean-field $M$-state Potts model\cite{PottsWu,mittag}. To describe the phase transition in the Pott model, one can define the complex order parameter\cite{PottsWu}
as
\begin{equation}
\phi=\sum_{\alpha=1}^M x_\alpha e^{\frac{2\pi i\alpha}{M}}\equiv \Phi e^{i\Theta}.\label{op}
\end{equation}
A real-value order parameter can also be defined as\cite{PottsWu,mittag} 
\begin{equation}s\equiv\frac{M   x_{\rm max}-1}{M-1} \label{s}\end{equation}
where $x_{\rm max}$ is the maximal of \{$\langle x_\alpha\rangle\}_{\alpha=0,1\cdots,M-1} $.

From the above mapping, the $M$ urns on a ring can be viewed as interacting Potts spin system with the spin state related by $Z_M$ discrete symmetry as depicted schematically in Fig. \ref{schematic}.
The magnetization along the $\alpha^{th}$ direction can be represented by a vector pointing in that direction with magnitude $x_\alpha$ as shown in Fig. \ref{schematic}a. The disordered state  refers to the situation that 
 the magnetization vectors in all $M$ directions are of equal magnitude which corresponds to the uniform state with equal particle fractions in each urn in the $M$-urn model.
 The transition of a particle to a neighboring urn can be represented by a ``tick of the clock" in Fig. \ref{schematic}a in the spin direction of a Potts spin.

Throughout this paper, we shall focus on the case of homogeneous ferromagnetic coupling case with $g_\alpha=g<0$ hereafter.
The first-order transition given in Eq. (\ref{gt01}) corresponds to the order(first non-uniform state)-disordered(uniform state) transition in the mean-field $M$-state Potts model which agrees with the results in Ref. \cite{mittag}. We note that the non-uniformity of the first non-uniform state is related to the order parameter of the Potts model $s$ in Eq. (\ref{s}) (which characterizes the order-disordered transition)\cite{PottsWu} by $\Psi^{(1)}=\sqrt{\frac{2}{M}} s$.

The various uniform or non-uniform equilibrium, non-equilibrium steady states discovered in the Ehrenfest urn models in Refs. \cite{cheng20} and \cite{cheng21}  can be conveniently represented graphically in terms of the Potts spin directions as depicted in Fig. \ref{schematic}b for the case of $M=3$. For the equilibrium case, the uniform state in the urn model (or the disordered state in the Potts model) can be represented by the $M$ symmetric magnetization vectors separated by $360\deg/M$ as depicted by the left panel in Fig. \ref{schematic}b. The non-uniform state (the order state in Potts model) is signified by one longer magnetization and $M-1$ shorter ones of equal lengths due to the spontaneous broken $Z_M$ symmetry as depicted by the middle panel in Fig. \ref{schematic}b. For the NESSs in the urn model (and the corresponding Potts model), the spin magnetization picture is the same as the uniform equilibrium case. On the other hand, for the non-uniform NESS with $p>q$, the symmetry of the $M-1$ shorter directions is further broken and has different lengths(see the right panel in Fig. \ref{schematic}b).

 \section{ Non-equilibrium  model of $M$ urns on a ring and the Potts spin Dynamics  }
 For the  urns/Potts model at equilibrium, detailed balance is obeyed and one possible  (microscopic) transition probability for  a spin  from the state $\alpha$ to state $\alpha'$ is given by the Glauber dynamics \cite{cheng21}:
 \begin{equation}
     T_{\alpha\to\alpha'}=\frac{1}{1+e^{\beta\Delta{\cal H}}}=\frac{1}{1+e^{\beta g(x_{\alpha'}-x_{\alpha})}},\label{Taa}
 \end{equation}
 and other spin updating dynamics is possible as long as the detailed balance is satisfied.
On the other hand, for non-equilibrium spin models, one needs to specify the dynamical or updating rules for the spins.  For example, there are different types of kinetic Ising models\cite{Dattagupta2004} using Glauber or Kawasaki dynamics in combination with sequential or parallel updating for the spins, which give different non-equilibrium properties.

The non-equilibrium dynamics in the urn model on a ring described in  Sec. 2 violates the detailed balance condition but still can be phrased in terms of the Potts spins and Hamiltonian in Eq. (\ref{PottsHam}). 
In particular, the system of  $M$ urns arranged in a ring with intra-urn interactions has recently been studied showing interesting non-equilibrium thermodynamics and phase transition behavior.
Here we will map the non-equilibrium $M$-urn model on a ring in Ref. \cite{cheng21} to a Potts model with appropriate spin transition dynamics. 
In this case, the flow of particles is restricted to neighboring urns on the ring and thus the only possible transitions are $  T_{\alpha\to\alpha\pm1}$. In terms of the Potts spins, 
the net transition rate for  spin $i$ to make the transition from state $\alpha$ to $\alpha+1$, as depicted schematically in Fig. \ref{schematic}a, is given by
\begin{equation}
     K^{(i)}_{\alpha\to\alpha+1}= p \delta_{\sigma_i\alpha}T_{\alpha\to\alpha+1}-q\delta_{\sigma_i\alpha+1} T_{\alpha+1\to\alpha}.
\end{equation}
Since the urn labels correspond to the Potts spin states, the  transition rates per particle for  urns in a 1D ring in Ref. \cite{cheng21} can be expressed in terms of the Potts spins as
\begin{eqnarray}
 W _{\sigma_k\to\sigma_k+n}&=& 0 \quad \hbox{if } n\neq\pm 1,\nonumber\\
  W _{\sigma_k\to\sigma_k+1}&=& \frac{p}{N}T_{\sigma_k\to\sigma_k+1}\sum_{i=1}^N\delta_{\sigma_i\sigma_k}=px_{\sigma_k}T_{\sigma_k\to\sigma_k+1} \label{W}\\
   W _{\sigma_k\to\sigma_k-1}&=& \frac{q}{N}T_{\sigma_k\to\sigma_k-1}\sum_{i=1}^N\delta_{\sigma_i\sigma_k}=qx_{\sigma_k}T_{\sigma_k\to\sigma_k-1},\nonumber
\end{eqnarray}
where it is understood that the value of the spin state is always under modulus $M$ to respect the cyclic representation of the Potts states (or the periodic boundary condition of the urns in a ring). For the special case of $p=q$, it reduces to the equilibrium case, and the detailed balance is obeyed. Thus the non-equilibrium urn model on a ring introduced in Ref. \cite{cheng21} can be viewed as a mean-field $M$-state Potts with special transition rates restricted to neighboring spin states as given by Eq. (\ref{W}) that breaks detailed balance in general and gives rise to a variety of non-equilibrium states.  The transition between the neighboring Potts spin states for spin $i$ can be viewed as a stochastic `ticking clock' in which the spin direction $\sigma_i$ undergoes clockwise or counter-clockwise discrete rotation of an angular step of $\frac{2\pi}{M}$ (see Fig. \ref{schematic}a).
The net transition rate  of the  mean magnetization from $\alpha\to\alpha+1$ of the system   is then
\begin{eqnarray}
    K_{\alpha\to\alpha+1}&=&\frac{1}{N} \sum_{i=1}^N K^{(i)}_{\alpha\to\alpha+1}= px_\alpha T_{\alpha\to\alpha+1}-qx_{\alpha+1} T_{\alpha+1\to\alpha}\label{K1}\\
    &=& W _{\alpha\to\alpha+1}-W _{\alpha+1\to\alpha}= \frac{ px_\alpha e^{gx_\alpha}-qx_{\alpha+1}e^{gx_\alpha+1}}{e^{gx_\alpha}+e^{gx_\alpha+1}},\label{K2}
\end{eqnarray}
which is also the mean particle flux from the $\alpha$ to $\alpha+1$ urns in the urn model\cite{cheng21}.

The NESSs in the urns model correspond to the steady flipping of the  Potts spin state $\alpha\to\alpha+1$ in a cyclic manner with the mean magnetizations remaining constant in time and that is why the uniform NESS and the uniform equilibrium state have the same magnetization diagram shown in the left panel of Fig. \ref{schematic}b.  
It is clear from the above discussion that the mean magnetizations are the same for the uniform equilibrium state and uniform NESS, thus to distinguish the non-equilibrium and equilibrium states in the Potts variables, one needs to measure the net mean transition rates of the spin directions from $\alpha\to \alpha+1$ by monitoring the microscopic transitions in the MC simulations. One can measure the number of transitions from $\sigma_\alpha\to \sigma_{\alpha+1}$ in a given period of time and average over all the spins as given by Eq. (\ref{K1}). One expects $\langle K_{\alpha\to \alpha+1} \rangle\simeq 0$ for equilibrium states,  whereas $|\langle K_{\alpha\to \alpha+1} \rangle| > 0$ for non-equilibrium states. 
In addition, one can also measure the time-dependent transition rates $K_{\alpha\to\alpha+1}(t)$ from the time-series data of $x_\alpha(t)$ using Eq.(\ref{K2}).

 \section{Monte Carlo Simulation results}
To explicitly verify the theoretical results in previous sections, one can carry out Monte Carlo simulations for the $M$-state Potts system. In such a simulation, a total of $N$ ($N$ is an integer multiple of $M$) Potts spins are placed in the system. The simulation algorithm starts by choosing 
a spin $i$ (whose state is $\sigma_i$) at random with uniform probability and the final state of the spin is also chosen randomly from the allowed final states with the transition rate given by Eq. (\ref{W}). In particular, for the urn model on a ring, the two allowed final states are chosen with probability $p$ for $\sigma_i+1$ and $q=1-p$ for $\sigma_i-1$ respectively.
Then the trial spin transition is made with probability given by Eq. (\ref{Taa}).
If $p=q=\frac{1}{2}$, then the transition rules satisfy the detailed balance condition and the system is at equilibrium. 
In general if $p>q$, there will be a counter-clockwise flux and a non-equilibrium state can be achieved.
After some sufficiently long transient time, the mean magnetization (or population fraction of an urn)  $x_\alpha(t)$, is measured using Eq. (\ref{xa}), and the complex order parameter ($\phi(t)$) in Eq. (\ref{op})  of the Potts spin system is monitored.

On the other hand, due to the equivalence of the mean-field Potts model and the urn model, one can also simulate directly the $M$-urn system on a network as was done in Ref. \cite{cheng20} for the equilibrium case. For simulations of the particles in the urns, the algorithm starts by choosing a particle at random with uniform probability, say  the chosen particle is in the $\alpha^{th}$ urn, then its jumping
probability to another urn $\alpha'$ is given by  the weighted transition $p_{\alpha\alpha'} T_{\alpha\to\alpha'}$, where $p_{\alpha\alpha'}$ is a normalized weighing factor and $T_{\alpha\to\alpha'}$ is given by Eq. (\ref{Taa}).
For the equilibrium case, there is no bias in the weighting factor, i.e. unweighted, and one can simply put  $p_{\alpha\alpha'}=1$. On the other hand, for the non-equilibrium model of $M$ urns on a ring, the allowed jumping probabilities are $pT_{\alpha\to\alpha+1}$  and $(1-p)T_{\alpha\to\alpha-1}$ respectively for the counter-clockwise and clockwise jumps respectively.

In this paper, we carry out MC simulations of the 3-state Potts model with the clockwise jumping rate $p$ and transition rates specified by Eq. (\ref{W}). Time evolution of the detailed dynamics of the order parameter is monitored and the probability distributions of the amplitude and phase of the Potts order parameter are measured.  Time is in the unit of Monte Carlo Steps per particle (MCS/N). One MCS/N means that on average every spin has attempted a flipping.
Most simulations are performed with $N=1500$
 spins, which is already large enough to demonstrate the predicted phases and their transitions (see Figs. \ref{MCeqm2} and \ref{MCNEQ}).
 
\subsection{Equilibrium first-order transition}
\begin{figure}[H]
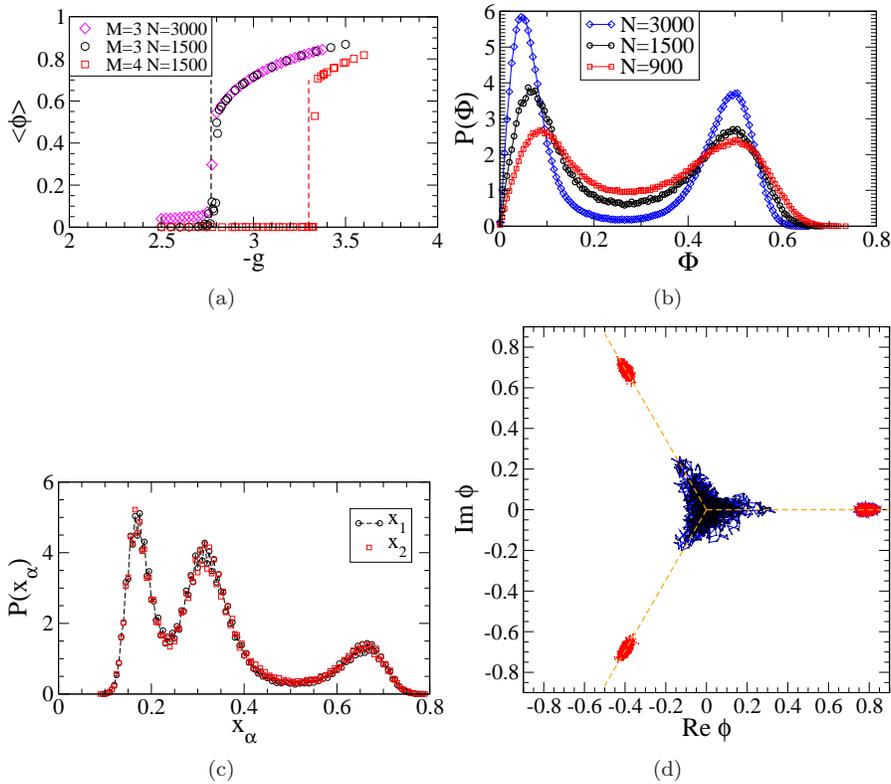

\centering
\subfigure[]{\includegraphics*[width=.48\columnwidth]{EqmPhivsgb.eps}} 
\subfigure[]{\includegraphics*[width=.48\columnwidth]{q3PPhib.eps}} 
\subfigure[]{\includegraphics*[width=.48\columnwidth]{q3EqmPx1x2.eps}}
\subfigure[]{\includegraphics*[width=.48\columnwidth]{q3Eqmphig2_73_2.eps}}
 \caption{Monte Carlo simulation results of the ferromagnetic Potts model at equilibrium for $N=1500$.  (a) The average order parameter amplitude, $\Phi$,  of the 3-state and 4-state Potts models plotted against the coupling $g$ (in the unit of $k_BT$). The vertical dashed lines are the theoretical first-order transition values given by Eq. (\ref{gt01}).  The result with $N=3000$ for the 3-state model is also shown. (b)  The measured distribution function of  $\Phi$, for the 3-state Potts model near the first-order transition at  $g=-2.77$. The co-existing phases are clearly revealed from the two peaks.  Results for $N=300$ and $N=3000$ are also displayed to show that the barrier between the coexisting phases increases with $N$. $10^6$ MCS/N are used to obtain good statistics. (c)  The measured distributions of the two directions of the magnetization, $P(x_\alpha)$ for the case in (b). (d)  Time evolution of the complex order parameter $\phi$ in its complex plane for a duration of 5000  MCS/N for $g=-2.7$ (black) and the three symmetry-related states of $g=-3.2$ (red) which are prepared separately with different initial conditions. The dashed lines mark the three Potts spin directions of 0 and $\pm\frac{2\pi}{3}$.} \label{MCeqm2}
\end{figure}

We first simulate the 3-state and 4-state Potts model at equilibrium (i.e. with equal jumping rates to other spin states) with the mean-field Hamiltonian in Eq. (\ref{PottsHam}). Fig. \ref{MCeqm2}a shows the average amplitude of the order parameter given by Eq. (\ref{op}) as a function of the coupling $g$ in the unit of $k_BT$. $\langle \Phi\rangle $ shows an abrupt jump as the coupling increases to a threshold value that agrees well with the theoretical first-order transition point in Eq. (\ref{gt01}), for both the 3-state and 4-state Potts model. The first-order transition nature is revealed clearly in the two coexisting peaks in the distribution function of $\Phi$ as shown in Fig. \ref{MCeqm2}b near the transition point for the   3-state Potts model. Results for smaller ($N=300$) and larger ($N=3000$) system sizes are also shown to confirm that the system size employed in the simulations is sufficiently large to observe the characteristic behavior of coexisting uniform and non-uniform equilibrium phases, and that the barrier between these coexisting phases increases with $N$. 
The distributions for two of the magnetizations near the transition point, which coincide with each other, display three peaks that correspond to the coexistence of the disordered states($x_\alpha\simeq \frac{1}{3}$)  and the ordered state, as shown in Fig. \ref{MCeqm2}c. The state of the system can be revealed in detail by examining the evolution of the order parameter $\phi(t)$ on its complex plane. Fig. \ref{MCeqm2}d  displays the time evolution of  $\phi(t)$  for a duration of 5000  MCS/N for $g=-2.7$ (the disordered state, black) and of $g=-3.2$ (the three symmetry-related disordered states, red). Each disordered state is prepared separately with a different initial condition, and the hopping between these three symmetry-related disordered states will take a very long time due to the high free energy barrier separating them. 

\subsection{NESSs and spin state transition rates}
\begin{figure}[H]
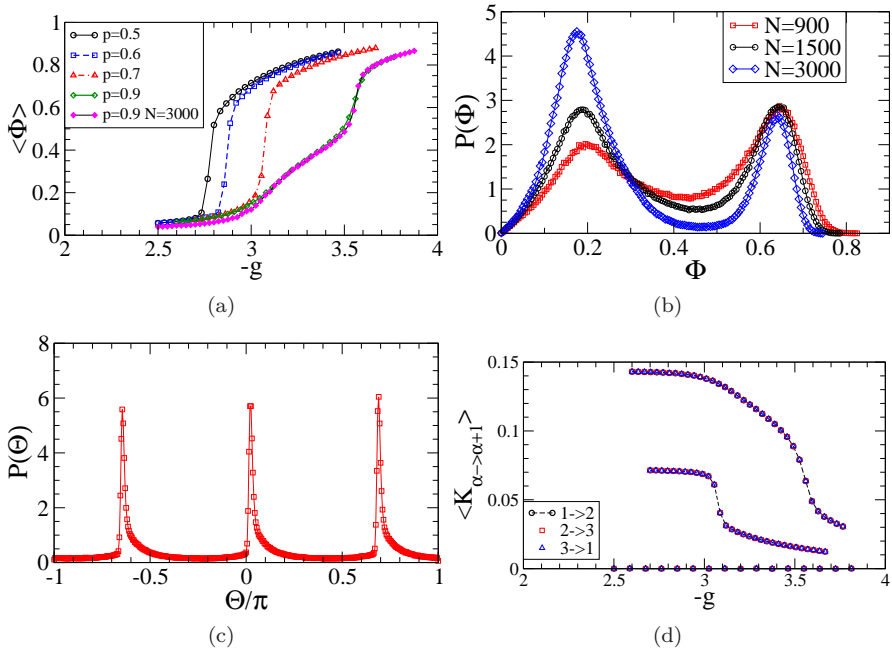

\centering
  \subfigure[]{\includegraphics*[width=.48\columnwidth]{q3PhivsgM3b.eps}}
      \subfigure[]{\includegraphics*[width=.48\columnwidth]{q3PPhip_7g3_07b.eps}}
      \subfigure[]{\includegraphics*[width=.48\columnwidth]{q3PThep_7g3_07.eps}}
      \subfigure[]{\includegraphics*[width=.48\columnwidth]{q3aveKvsgp_9.eps}}
 \caption{Monte Carlo simulation results of the non-equilibrium ferromagnetic 3-state Potts model with $p>\frac{1}{2}$. $N=1500$.  (a) The average amplitude of the order parameter, $\langle \Phi\rangle$, is plotted against the coupling $g$ for different values of $p$. System size of $N=3000$ is also shown for the $p=0.9$ case. (b)  The measured distribution function of $\Phi$ for $p=0.7$ near the non-equilibrium first-order transition at  $g=-3.07$. The co-existing phases are clearly revealed from the two peaks. (c)   The corresponding distribution function of $\Theta$ for the case in (b).  (d) The measured microscopic transition rates of the spin directions from $\alpha\to \alpha+1$  for the $p=0.5$ (equilibrium case, bottom line),  $p=0.7$ (middle curve), and  $p=0.9$ (upper curve) cases.  } \label{MCNEQ}
\end{figure}

Here we consider the 3-state Potts model with asymmetric jumping rates for $\alpha\to \alpha+1$ (with jumping rate $p$)  and $\alpha\to \alpha-1$ (with jumping rate $q=1-p\neq p$), as depicted in Fig. \ref{schematic}a.
Fig. \ref{MCNEQ}a shows the average amplitude of the order parameter given by Eq. (\ref{op}) as a function of  $g$ (in the unit of $k_BT$). For smaller values of $p$, $\langle \Phi\rangle $ shows an abrupt jump as the coupling increases at some threshold value, signaling a first-order transition similar to the equilibrium ($p=\frac{1}{2}$) case. But for larger values of $p$ (see the $p=0.9$ curve), the sharp jump is replaced by a gradual increase in  $\langle \Phi\rangle $ for $-g$ between 3 and $\sim 3.5$, then followed by a steeper increase, suggesting a more complex non-equilibrium dynamics.  The first-order non-equilibrium phase transition is revealed from the two peaks of $P(\Phi)$ at  $g=-3.07$ for $p=0.7$, as shown in Fig. \ref{MCNEQ}b. 
Results for smaller ($N=300$) and larger ($N=3000$) system sizes are also shown to confirm that the system size employed in the simulations is sufficiently large to observe the characteristic behavior of coexistence of the two NESS phases, and that the barrier between them increases with $N$. The relative heights of the two peaks vary with the system sizes arising from the systematic shift of the phase transition thresholds resulting from the finite size effects, which is also present for the equilibrium phase transition as shown in Fig. \ref{MCeqm2}b.
The associated distribution for the phase of the order parameter is plotted in Fig. \ref{MCNEQ}c, showing that $\phi$ points along the three Potts directions.
For smaller values of $p$, the system is in a NESS and its nature cannot be revealed by merely measuring the steady-state distributions such as $P(\Phi)$ and $P(\Theta)$ as they show similar features as the equilibrium case. As suggested by the result of the 3-urn model in Ref. \cite{cheng21}, the equilibrium state and NESS differ in the non-vanishing mean particle flux (which corresponds to the microscopic spin direction transition rate for the Potts spin model). The microscopic transition rates of the spin directions from $\alpha\to \alpha+1$, $\langle K_{\alpha\to \alpha+1}\rangle$, are measured in the simulations and the results are shown in Fig. \ref{MCNEQ}d for the equilibrium and non-equilibrium Potts models. It is clear that $\langle K_{\alpha\to \alpha+1}\rangle$ vanishes for the equilibrium case while  $\langle K_{\alpha\to \alpha+1}\rangle$ is significantly positive for the cases of $p=0.7$ and 0.9. In addition,  $\langle K_{\alpha\to \alpha+1}\rangle$ shows a sharp change at some value of $g$, indicating the occurrence of a non-equilibrium phase transition. 

\begin{figure}[H]
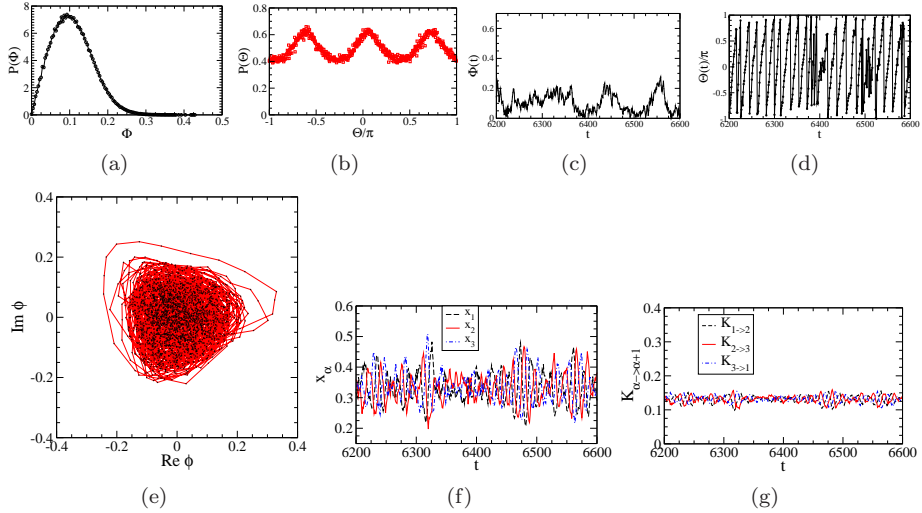

\centering
   \subfigure[]{\includegraphics*[width=.243\columnwidth]{q3PPhip_9g2_9.eps}}
      \subfigure[]{\includegraphics*[width=.243\columnwidth]{q3PThep_9g2_9.eps}}
      \subfigure[]{\includegraphics*[width=.243\columnwidth]{q3Phitp_9g2_9.eps}}
    \subfigure[]{\includegraphics*[width=.243\columnwidth]{q3Thetatp_9g2_9.eps}}
      \subfigure[]{\includegraphics*[width=.325\columnwidth]{q3phip_9g2_9.eps}}  
        \subfigure[]{\includegraphics*[width=.325\columnwidth]{q3xtp_9g2_9.eps}}
        \subfigure[]{\includegraphics*[width=.325\columnwidth]{q3Kp_9g2_9.eps}} 
  \caption{MC simulations for the uniform  NESS in the 3-state Potts model with $p=0.9$ and $g=-2.9$, $N=1500$.   (a)  The measured distribution function of $\Phi$.  (b)   The corresponding distribution function of $\Theta$ for the case in (a). (c) Time course of the amplitude of the order parameter. $\Phi(t)$,  showing the stochastic bursts of the amplitude. (d) Time courses of the phase $\Theta(t)$ in (c), showing the oscillatory feature during the amplitude burst.  Time is in the unit of Monte Carlo Steps per spin (MCS/N). (e) Time evolution of the complex order parameter $\phi$ in its complex plane. (f)  Time evolution of the Potts spin magnetizations $x_\alpha(t)$. (g) Time evolution of the mean transition rates of the Potts spin directions as measured using Eq. (\ref{K2}).
  } \label{uNESS}
\end{figure}

As $p$ becomes large, the system is far from equilibrium and the net transition rate of mean magnetizations from $\alpha\to \alpha+1$ becomes larger. We first examine the case when the ferromagnetic coupling is not so strong and the system is still in a disordered state or the uniform NESS (the fraction of particles in the urns is uniform) as discussed in Ref. \cite{cheng21} in terms of the urn model.
MC simulation results for the uniform  NESS in the 3-state Potts model with $p=0.9$ and $g=-2.9$ are shown in Fig.  \ref{uNESS}. The disordered nature of the uniform state can be seen in the single peak at a small value of $\Phi$ in the distribution of the amplitude of the order parameter shown in  Fig.  \ref{uNESS}a. The phase of the order parameter shows broad distribution peaks slightly ahead of the three Potts spin directions of 0 and $\pm\frac{2\pi}{3}$ as shown in  Fig.  \ref{uNESS}b. More details about the uniform NESS state can be seen from the time traces of  $\Phi(t)$ and $\Theta(t)$ shown respectively in  Figs.  \ref{uNESS}c and   \ref{uNESS}d.
$\Phi(t)$  displays   stochastic bursts accompanied with the oscillatory features in $\Theta(t)$. The trajectory of $\phi(t)$ in  Fig.  \ref{uNESS}e reveals that the order parameter fluctuates around with very small magnitude most of the time with occasional stochastic periodic counter-clockwise excursions. The time traces of the mean magnetization in the three Potts directions are shown in Fig.  \ref{uNESS}f, indicating that the system steadily fluctuates around the disordered (uniform) NESS. The key feature signifying the uniform NESS nature can be seen in Fig. \ref{uNESS}g for the time traces of the spin transition rates measured using Eq. (\ref{K2}), which shows that all three net transitions fluctuate about the same finite mean value.

\begin{figure}[H]
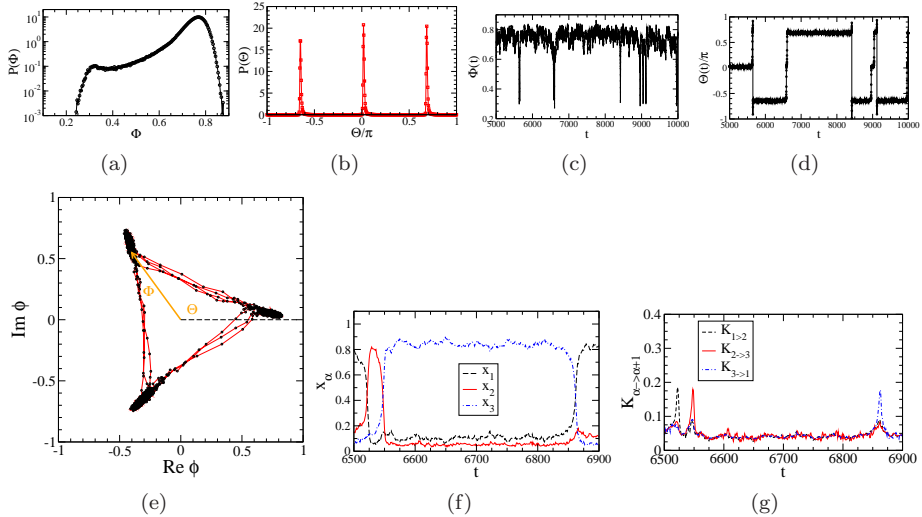

\centering
     \subfigure[]{\includegraphics*[width=.243\columnwidth]{q3PPhip_9g3_6.eps}}
    \subfigure[]{\includegraphics*[width=.243\columnwidth]{q3PThep_9g3_6.eps}}   \subfigure[]{\includegraphics*[width=.243\columnwidth]{q3Phitp_9g3_6.eps}}
     \subfigure[]{\includegraphics*[width=.243\columnwidth]{q3Thetap_9g3_6.eps}}
  \subfigure[]{\includegraphics*[width=.325\columnwidth]{q3phip_9g3_6new.eps}}
   \subfigure[]{\includegraphics*[width=.325\columnwidth]{q3xtp_9g3_6.eps}} 
    \subfigure[]{\includegraphics*[width=.325\columnwidth]{q3Kp_9g3_6.eps}} 
         \caption{ MC simulations for the non-uniform NESS with $p=0.9$ and $g=-3.6$, $N=1500$.  (a)  The measured distribution function of $\Phi$. (b)   The corresponding distribution function of $\Theta$ for the case in (a). (c) Time courses of the amplitude $\Phi(t)$, and (d) phase $\Theta(t)$ of the order parameter.  Time is in the unit of Monte Carlo Steps per particle (MCS/N). (e) Time evolution of the complex order parameter $\phi$ in its complex plane. The amplitude $\Phi$ and phase $\Theta$ of the complex order parameter $\phi$ are also shown schematically. (f)  Time evolution of the Potts spin magnetizations $x_\alpha(t)$.  (g)  Time evolution of the mean transition rates of the Potts spin directions as measured using Eq. (\ref{K2}).  } \label{nuNESS}
\end{figure}

As the ferromagnetic coupling grows, the NESS undergoes a first-order transition from a disordered (uniform) NESS to an ordered (nonuniform) NESS. Fig. \ref{nuNESS} shows the MC simulation results for the non-uniform  NESS  with $p=0.9$ and $g=-3.6$.  The distribution of the amplitude of the order parameter in  Fig. \ref{nuNESS}a displays a major peak at $\Phi\lesssim 1$ corresponding to the order (non-uniform) NESS. The transient states for the transitions between the three symmetry-related NESSs can be seen from the shoulder at small values of $\Phi$ revealed in the semi-log plot.  The corresponding distribution of $\Theta$ is plotted in Fig. \ref{nuNESS}b, showing three sharp peaks slightly ahead of the three Potts directions. The peaks in $P(\Theta)$ are much sharper as compared to the uniform NESS indicating that the directional fluctuations of Potts spins are much smaller for the non-uniform NESS due to the strong ferromagnetic coupling. Details for the non-uniform NESS state can be seen from the time traces of  $\Phi(t)$ and $\Theta(t)$ shown respectively in  Figs.  \ref{nuNESS}c and   \ref{nuNESS}d.  $\Phi(t)$ fluctuates about some large mean value with fast stochastic dips that occur occasionally.  $\Theta(t)$ reveals that these stochastic dips in $\Phi$ are due to the collective global flipping of the spins from $\alpha \to\alpha+1$ as shown in Fig. \ref{nuNESS}d. The above picture for the non-uniform NESS  is confirmed by the trajectory of $\phi(t)$ shown in  Fig.  \ref{nuNESS}e, indicating that the order parameter points at one of three symmetry-related Potts directions with some fluctuations for most of the time, and   $\phi$ flips one step counter-clockwise stochastically. Such a counter-clockwise stochastic global flipping of the spins does not occur frequently due to the free energy barrier separating the three symmetric non-uniform NESSs. By comparing with the non-uniform state at equilibrium in Sec. 5.1 in which such a stochastic global flipping of the spin was never observed in similar simulation durations, one can deduce that the free energy barrier separating the three symmetric non-uniform states is significantly lower for the NESS case.
The time traces of the mean magnetization in the three Potts directions are displayed in Fig.  \ref{nuNESS}f showing that the system predominantly fluctuates steadily around the ordered (non-uniform) NESS,  with occasional stochastic global flipping of the spins. The  NESS nature of the non-uniform state can be seen in the time traces of the spin transition rates shown in Fig. \ref{nuNESS}g.  All three net transitions fluctuate about the same finite mean value which is somewhat lower than that of the uniform NESS, apart from the occasional stochastic bursts due to the global flipping of the spins between the
three symmetry-related Potts spin states.

\subsection{NEPS and spin rotation}
\begin{figure}[htb]
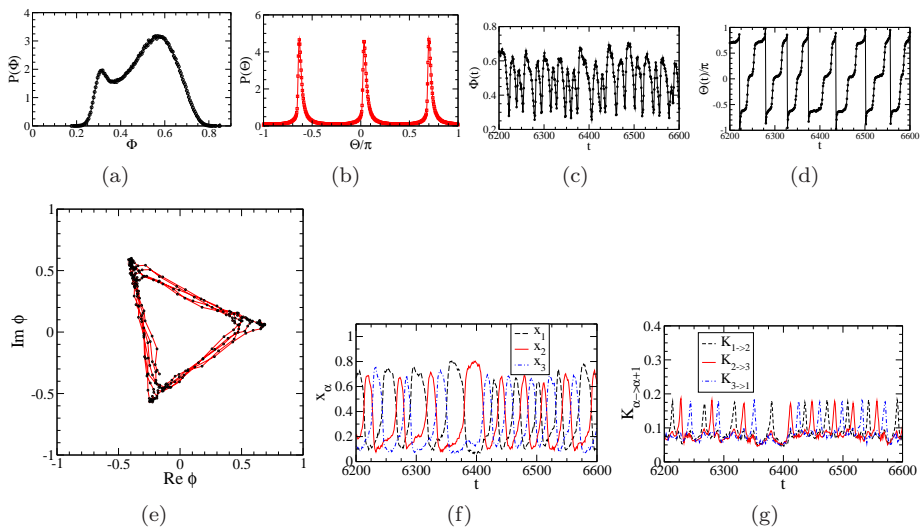

\centering
  \subfigure[]{\includegraphics*[width=.2453\columnwidth]{q3PPhip_9g3_5.eps}}
      \subfigure[]{\includegraphics*[width=.243\columnwidth]{q3PThep_9g3_5.eps}}  \subfigure[]{\includegraphics*[width=.243\columnwidth]{q3Phitp_9g3_5.eps}}
     \subfigure[]{\includegraphics*[width=.243\columnwidth]{q3Thetatp_9g3_5.eps}}
      \subfigure[]{\includegraphics*[width=.325\columnwidth]{q3phip_9g3_5.eps}}
       \subfigure[]{\includegraphics*[width=.325\columnwidth]{q3xtp_9g3_5.eps}}
       \subfigure[]{\includegraphics*[width=.325\columnwidth]{q3Kp_9g3_5.eps}} 
         \caption{MC simulations for the NEPS with $p=0.9$ and $g=-3.5$. $N=1500$.  (a)  The measured distribution function of $\Phi$. The co-existing phases are revealed from the two peaks. (b)   The corresponding distribution function of $\Theta$ for the case in (a). (c) Time courses of the amplitude $\Phi(t)$, and (d) phase $\Theta(t)$ of the order parameter.  Time is in the unit of Monte Carlo Steps per particle (MCS/N). (e) Time evolution of the complex order parameter $\phi$ in its complex plane.  $\phi$ undergoes a continuous counter-clockwise rotation with non-uniform angular speed in its complex plane. (f)  Time evolution of the Potts spin magnetizations $x_\alpha(t)$. (g) Time evolution of the mean transition rates of the Potts spin directions as measured using Eq. (\ref{K2}).} \label{NEPS}
\end{figure}

For sufficiently large values of $p$, a new non-steady non-equilibrium state is observed in the non-equilibrium Potts model that was not reported in detail in the corresponding urn model in Ref. \cite{cheng21}.
Such a non-steady non-equilibrium state occurs in some intermediate strength of ferromagnetic coupling and undergoes periodic oscillation masked by thermal fluctuation, and is termed non-equilibrium period state (NEPS). 
Fig. \ref{NEPS} shows the MC simulation results for the NEPS  with $p=0.9$ and $g=-3.5$.  The distribution of the amplitude of the order parameter in  Fig. \ref{NEPS}a displays a major peak at larger $\Phi$ and a smaller peak at lower  $\Phi$. Given the kink in the $\langle \Phi\rangle$ vs. $-g$ curve in Fig. \ref{MCNEQ}a, our result suggests there is a first-order phase transition between the non-uniform NESS and NEPS. 
The order parameter $\phi(t)$ undergoes continuous rotation on its complex plane with a non-uniform angular velocity, slowing down around the three major Potts spin directions,  as depicted in Fig. \ref{NEPS}b.  The corresponding distribution of $\Theta$ is plotted in Fig. \ref{NEPS}b showing three peaks slightly ahead of the three Potts directions. The peaks in $P(\Theta)$ are less sharp as compared to the non-uniform NESS (see Fig. \ref{nuNESS}b). The periodic nature of NEPS cannot be seen merely from the result of the $P(\Phi)$ and $P(\Theta)$ distributions, but instead one needs to examine in detail the time traces of $\Phi$ and $\Theta$ which are displayed in Fig.  \ref{NEPS}d. 
$\Phi(t)$ shows oscillations with fluctuating amplitudes accompanied by small fluctuations in the period as shown in Fig.  \ref{NEPS}c. The clear periodic oscillations in $\Theta(t)$  reveal the periodic nature of the NEPS along with the continuous counter-clockwise rotation of the order parameter. The periodic counter-clockwise rotation with a non-uniform angular speed being slowed down around the three Potts directions is also confirmed in the trajectory of $\phi(t)$ shown in Fig.  \ref{NEPS}e. The time traces of the mean magnetization in the three Potts directions are displayed in Fig.  \ref{NEPS}f, showing periodic phase-locked oscillations between the three magnetizations (or the phase-locked periodic oscillations of particle fractions in the three urns).  Finally, the periodic dynamics also show up in the time traces of the spin transition rates shown in Fig. \ref{NEPS}g, displaying the same phase-locked periodic oscillations for the three transition rates  (or the phase-locked periodic oscillations of particle fluxes between three urns).

  \section{Summary and Outlook }
In this paper,  we established the equivalence of the multi-urn Ehrenfest model with intra-run interactions and the Potts spin model for the equilibrium and non-equilibrium cases. The non-equilibrium dynamics are introduced with bias jumping rates for the clockwise and counter-clockwise spin transition directions. Such a bias jumping rate, specified by the parameter $p$ in the current model, might appear ad hoc at first sight, but as shown in Ref. \cite{cheng21} it can be interpreted in terms of an effective chemical potential $\mu=k_BT\ln(\frac{p}{q})$ that drives the Potts spin direction from $\alpha\to \alpha+1$. 
 Such a driving can possibly be realized by a rotating magnetic field. 
Similar to the urn model, there are non-trivial non-equilibrium phase transitions for the non-equilibrium Potts spin model, reflecting different levels of non-equilibria. Monte Carlo simulation of the Potts spin model is performed to investigate and confirm that the equilibrium and non-equilibrium phase transitions observed in the Ehrenfest multi-urn model are also observed in the equivalent Potts spin model. In particular, there are interesting non-equilibrium phase transitions between the disordered and ordered Potts NESSs. In addition, new
NEPS exists for the far-from-equilibrium case as shown in the simulations of the non-equilibrium Potts model, whose detailed mechanism for its occurrence can be further investigated theoretically in the framework of the $M$-urn model in terms of bifurcation theory in nonlinear dynamics.

 For higher Potts states with $M\geq4$, due to the complexity in the underlying nonlinear dynamics, it is possible that the system could display very far-from-equilibrium dynamics such as quasi-periodic or chaotic non-equilibrium states. The $M\geq4$ cases are under investigation and the results will be reported in the future. Furthermore, the associated entropy production in these non-equilibrium states with different degrees of non-equilibrium is another interesting issue, which can provide valuable insight and understanding of how the energetics and dissipation are being transported.

   \section*{Acknowledgement}  This work has been supported by the National Science and Technology Council of Taiwan under grants Nos. 110-2112-M-008-026-MY3 and 111-2112-M-018-005.

\bibliography{references}

\end{document}